\documentclass[12pt,oneside]{article}

\usepackage{amsmath, amssymb, fullpage, mathrsfs, graphicx}
\usepackage[usenames, dvipsnames]{color}
\usepackage{float}
\usepackage[utf8]{inputenc}
\usepackage[marginparwidth=1.5in]{geometry}

\usepackage[backend=biber, sorting=none, style=numeric-comp]{biblatex}

\usepackage[usenames,dvipsnames]{xcolor} 
\usepackage{hyperref}
\hypersetup{
    colorlinks=true,       % false: boxed links; true: colored links
    linkcolor=MidnightBlue,%BrickRed,          % color of internal links
    citecolor=OliveGreen,        % color of links to bibliography
    filecolor=magenta,      % color of file links
    urlcolor=MidnightBlue           % color of external links
}

\newcommand{\be}{\begin{equation}}
\newcommand{\ee}{\end{equation}}
\newcommand{\bea}{\begin{eqnarray}}
\newcommand{\eea}{\end{eqnarray}}

\bibliography{references}

\newcommand{\bL}{{\textbf L}}
\newcommand{\bM}{{\textbf E}}

\newcommand{\ba}{{\boldsymbol\alpha}}
\newcommand{\bb}{{\boldsymbol\beta}}

\newcommand{\bth}{{\boldsymbol \theta}}

\newcommand{\const}{\text{const.}}

\newcommand{\rd}{{\mathrm{d}}}

\newcommand{\scri}{{\mathscr{I}}}
\newcommand{\p}{{\partial}}
\newcommand{\po}{\partial_\Omega}

\newcommand{\hg}{{\sqrt{\hat g}}}
\newcommand{\q}{{\sqrt q}}

\renewcommand{\d}{\delta}

\newcommand{\A}{{\mathcal A}}
\newcommand{\J}{{\mathcal J}}
\newcommand{\N}{{\mathcal N}}
\newcommand{\Dk}[1]{{\Delta^\ell_{#1}}}

\renewcommand{\O}{{\Omega}}
\renewcommand{\cite}{\parencite}

\newcommand{\seq}{{\;\overset{\scri}{=}\;}}
\newcommand{\oeq}{\hat{=}}
\title{ Asymptotic Renormalization in Flat Space:\\
 Symplectic Potential and Charges of Electromagnetism }

\author{Laurent Freidel\footnote{lfreidel@perimeterinstitute.ca}, \, Florian Hopfm\"uller\footnote{fhopfmueller@perimeterinstitute.ca}, \, and Aldo Riello\footnote{ariello@perimeterinstitute.ca}}

\date{\today}

\begin{document}

	\maketitle	

	{ We present a systematic procedure to renormalize the symplectic potential of the electromagnetic field at null infinity in Minkowski space. We work in $D\geq6$ spacetime dimensions as a toy model of General Relativity in $D\geq4$ dimensions. Total variation counterterms as well as corner counterterms are both subtracted from the symplectic potential to make it finite. These counterterms affect respectively the action functional and the Hamiltonian symmetry generators. The counterterms are local and universal. We analyze the asymptotic equations of motion and identify the free data associated with the renormalized canonical structure along a null characteristic. This allows the construction of the asymptotic  renormalized  charges whose Ward identity gives the QED soft theorem, supporting the physical viability of the renormalization procedure. We touch upon how to extend our analysis to the presence of logarithmic anomalies, { and upon how our procedure compares to holographic renormalization}.
}

{
\newpage
\small
\tableofcontents 
\normalsize
\newpage

%===========================================================

\section{Introduction}

The canonical formalism has proven very useful in the treatment of asymptotic symmetries and conservation laws of gauge theories on Minkowski space as well as asymptotically flat Einstein-Hilbert gravity. Asymptotic symmetries and conservation laws, in turn, have far-reaching connections to diverse topics such as scattering theory, soft gauge boson theorems, infrared dressings, and memory effects (see \cite{Strominger:2017zoo} for an overview).

In the presence of boundaries, the complete algebra of gauge generators include non-vanishing charges for those gauge transformations that do not vanish at the boundary. This is the case both at finite boundaries \cite{Balachandran:1994up, Carlip:1996mi} and at conformal infinity \cite{Regge:1974zd, He2014}.\footnote{The situation is enriched by the fact that dual charges have been recently suggested to also play a crucial role, once again both at asymptotic \cite{Strominger:2015bla, Hamada:2017bgi, Campiglia:2018see} and finite boundaries \cite{Hosseinzadeh:2018dkh, Freidel:2018fsk}, as well as in gravity \cite{Godazgar:2018qpq}.} 
 
 A natural expectation is that the asymptotic generators must arise as limits of the finite-boundary ones. However, the naive implementation of this idea encounters a fundamental difficulty: in higher dimensional gauge theories and in gravity, the symplectic potential (SP) -- that is the fundamental object required to construct the appropriate generators -- is found to diverge when the boundary is pushed to infinity (e.g. \cite{Compere:2018ylh}). { A possible solution is to impose additional boundary conditions to obtain a finite result in the limit.}  The problem is that this amounts to the exclusion of certain modes from the boundary phase space and this automatically results in a restriction of the algebra of the boundary generators. Hence, following this logic, one finds that the asymptotic generator algebra is \textit{smaller} than the finite boundary one -- which is an unsatisfactory conclusion.

One central example of this concerns the attempts to include super-rotations to the asymptotic symmetries of gravity. These are naturally part of the algebra at finite boundaries \cite{Donnelly:2016auv}, but are excluded in the standard construction of the asymptotic symmetries of gravity which only features the BMS group (see \cite{Ashtekar:2018lor} for an enlightening review, and references therein). In recent years, two different attempts at including super-rotations in the asymptotic symmetry structure have been made \cite{Barnich:2010eb, Barnich:2011mi, Campiglia:2014yka, Campiglia:2015yka} (see also  \cite{Kapec:2014opa, Distler:2018rwu} for the connection to the subleading soft graviton theorems). However, serious challenges in the definition of the super-rotation charges emerge \cite{Flanagan:2015pxa, Compere:2018ylh}.

To resolve these issues, and allow for a full enlargement of the symmetry algebra at infinity, we propose a  renormalization scheme, { that we name \textit{asymptotic renormalization}}, for the symplectic potential and the boundary charges. 
{ The crucial and only input necessary to run our renormalization scheme is the existence of a conformal compactification \`a la Penrose of the spacetime and the fields inhabiting it. 
 Echoing Penrose's language, we refer to the ensuing mathematical conditions on the electromagnetic field as ``asymptotic Maxwell conditions". 
 
The details of the renormalization scheme -- although not its viability, which is general -- depend on the way asymptotic infinity is approached.  
In the present paper, in view of a generalization to general relativity in asymptotically flat spacetimes, we work in Minkowski space and, in a neighbourhood of $\scri^+$, we adopt Bondi-like (i.e. retarded time) coordinates. 
This means that ``radial evolution'' happens along null characteristics, an important feature that markedly distinguishes the present framework from standard holographic renormalization \cite{skenderis2002lecture}.
In particular, some counterterms might depend on the dynamical field content.
This fact can be traced back to the implementation of a null radial evolution. 
However, this fact does not compromise the viability of the renormalization scheme which -- as we will discuss in a moment -- only exploits the cohomological ambiguities intrinsic to the symplectic potential. 

Our asymptotic renormalization procedure works in two steps: it first establishes a radial-evolution equation for the (pre)symplectic potential which encodes its renormalizability up to terms in the field-space and spacetime cohomology, and then -- by studying the equations of motion and in particular their radial evolution -- reexpresses the renormalized symplectic potential and the counterterms in terms of the free data intrinsic to $\scri$ and the free radiative data.
It is this second step that has widely different properties depending on whether the radial evolution happens  along a null or spacelike direction.
We believe that the null evolution is more universal and physically better motivated, even though its features can at first appear unsettling from a holographic renormalization perspective.

Nonetheless, were we to work in an asymptotically AdS space with spacelike radial evolution, we expect (but we do not prove) our scheme to be compatible with the standard holographic renormalization setup. 
The holographic renormalization program (see e.g.\cite{skenderis2002lecture}), which is usually focused on field theories on an AdS spacetime or on asymptotically AdS gravity, rather than gauge theory on Minkowski space, partly motivates the procedure devised here. 
Holographic renormalization usually focuses only on the renormalization of the action, by the addition of a {\it local counterterms intrinsic to the boundary}.\footnote{ Once again, it is this property of the counterterm that might fail in our asymptotic renormalization prescription when compared to the AdS holographic renormalization.}
Our work can be viewed as an extension of this idea where not only the action is renormalized, but also the symplectic structure via the addition of corner terms. As a result, we get a renormalization of the asymptotic charges.
 Papadimitriou \cite{papadimitriou2010holographic,Papadimitriou:2016yit} already pointed out the usefulness of extending holographic renormalization ideas to the Hamiltonian framework and to theories which do not necessarily admit a holographic dual. 
 The asymptotic renormalization of conserved charges for global symmetry such as the energy has already been considered in AdS \cite{BalasubramanianKraus1999,Hollands:2005ya,Compere:2008us}. Also,  holographic renormalization of asymptotically flat general relativity  has also been considered from a different perspective in \cite{Mann:2005yr, Park:2012bv}.
We leave a detailed comparison of our asymptotic renormalization with the holographic renormalization to future work .
}

{ In this paper, we illustrate the asymptotic renormalization scheme} in  the context of electromagnetism in higher spacetime dimensions ($D\geq6$).
Our methodology is designed to apply to general relativity in $D\geq4$ { and with any sign of the cosmological constant}.
Investigation of the latter case is in fact our original motivation, of which higher dimensional electromagnetism constitutes a toy model, and it will be treated in a forthcoming publication.
 
 The main reason such an asymptotic renormalization scheme is possible is that the  SP is inherently ambiguous \cite{Jacobson:1993vj}, in two separate ways. Firstly, altering the action by a boundary term adds a total variation term to the SP. This does not change the symplectic form and hence leaves the canonical theory and the generators of gauge transformations unchanged. 
Secondly, a total derivative can be added to the SP, which, when the SP is integrated on a portion of the asymptotic boundary of spacetime, becomes an integral on the codimension-2 boundaries of that portion. This corresponds to revising the corner phase space and modifying the generators of asymptotic gauge transformations.

 Some of the ideas we present here have been explored very recently in the context of 4D gravity by Compere et al. \cite{Compere:2018ylh}.
 These authors realized that if one wants to extend the symmetry transformations to include the super-boosts, it is necessary to consider terms that would make the naive SP divergent. They also show that a renormalized SP and renormalized super-boosts charges could be defined by the addition of corner terms and consider the corresponding memory effects.  Since the structure of divergences in 6D electromagnetism is similar to that of 4D general relativity, our work can be viewed as a { toy-model for a } covariant extension of their results in the context of the Maxwell field. 
 This extension allows the inclusion of  the full boundary algebra.  It also allows for a more geometric understanding of the renormalization procedure that will be extended to general relativity in future work.

The reason for the analogy between 6D electromagnetism and 4D general relativity is that, in both cases, the spacetime dimension is the dimension in which the theory is conformal plus $2$.
Throughout most of the paper, we will work in an arbitrary dimension $D\geq6$ because this brings out the structure of the asymptotic degrees of freedom more clearly.
We find that higher dimensional classical canonical electromagnetism on flat spacetimes is interesting in its own right, but has not to date received much attention from the asymptotic viewpoint, with the notable exception of \cite{Kapec:2014zla}.

The first part of the paper is very general, it applies to all dimensions, even or odd and it is also valid when the asymptotic expansion of the fields develop certain logarithms. We show how to renormalize the symplectic potential into a finite SP by the addition of corner and boundary terms.
This requires, in general, the introduction of conformally anomalous counterterms.
We also show that the resulting SP is layered into several canonical components that include, but are not limited to, a radiative pair, a Coulombic pair, and a soft pair. We also show that the corresponding asymptotic charges are finite.

In the second part of the paper, we analyze in detail the asymptotic equations of motion, which are needed to resolve the dependencies between the different canonical layers { along a {\it null} characteristic}.
We restrict our analysis to even dimensions and assume that the anomalous logarithmic terms all vanish. That is we restrict our analysis to asymptotically analytic solutions with no logarithm dependence in the radial direction.
This is a restriction which is always made in the usual asymptotic analysis (for a notable exception, see \cite{Herdegen:2016bio}).
In our context, we can understand precisely what this restriction entails.
Finally, we show that our definition of the renormalized charge is equipped with a conservation law perfectly compatible with the soft theorems derived in  \cite{Kapec:2014zla} (see also \cite{He2014}).

The structure of the paper is as follows: in section \ref{sec:spacetime}, we introduce the auxiliary compactified spacetime we work in.  In section \ref{sec:eomI}, we give our fall-offs conditions and implement the most divergent order of the equations of motion. In the core section \ref{sec:renormalize}, we renormalize the symplectic potential current. In section \ref{sec:generators}, we derive the gauge generators of the renormalized symplectic structure. In section \ref{sec:EoMII}, we analyze in detail the asymptotic equations of motion, provide the definition of the charge aspects and analyze the anomaly-freeness conditions. In section \ref{sec:sp7}, we construct the canonical pairs and analyze the soft conservation equations together with their relationship with the soft theorems. We conclude in section \ref{sec:conclusions}.

%================================================

\section{Spacetime structure}\label{sec:spacetime}

This section lays out the basic spacetime structures used in the remainder. We consider vacuum Maxwell theory on Minkowski spacetime of spacetime dimension $D\geq 5$.
One of the motivation is that the asymptotic structure and divergences
appearing in Maxwell theory for  $D=6$ are similar to the ones for gravity in $D=4$.

We will find useful to work in an auxiliary spacetime, the ``conformal frame'', obtained through a conformal compactification of Minkowski spacetime \`a la Penrose \cite{Penrose:1962ij, Penrose:1987uia,Penrose:1986ca}.  
The key advantage of this approach is that asymptotic infinity presents itself as a finite boundary in the conformal frame. 
The structure of infinity is reflected in the behavior of the conformally rescaled fields near this boundary. 
To avoid technicalities, we restrict our analysis to a coordinate patch, that of ``retarded Bondi coordinates'', which covers only future null infinity.

The Minkowski metric in retarded Bondi coordinates, with $u=t-r$, reads $\hat g_{ab} \rd x^a \rd x^b = - \rd u^2 - 2\rd u \rd r  + r^2 q_{AB} \rd x^A \rd x^B$, $q_{AB}$ being the metric of a unit round $(D-2)$-sphere. We introduce the coordinate $\O = 1/r$, and work in the conformally compactified spacetime with the rescaled metric $g_{ab} := \O^2 \hat g_{ab}$ and inverse $g^{ab}=\O^{-2} \hat g^{ab}$:
\begin{subequations}
\begin{align}
 g_{ab} \rd x^a \rd x^b 
 =& - \O^2 \rd u^2 + 2\rd u \rd \O  + q_{AB} \rd x^A \rd x^B\\
 g^{ab} \p_a \p_b 
 = &  \O^2 \po^2 + 2 \po \p_u + q^{AB} \p_A \p_B.
\end{align}
\end{subequations}
All indices will be contracted with respect to this metric unless otherwise specified by the use of hats, $\hat\cdot$.
Note that in these coordinates $\sqrt{|g|} = \q = \O^D \hg$.
In these coordinates, future null infinity $\scri^+\cong S^{D-2}\times\mathbb R$  (we will drop the plus in the following) is located at $\Omega=0$, corresponding to the null limit at $r \rightarrow \infty$ of the timelike level surfaces of $r$.
We define the conormal to the surfaces at $\O=\const$,
\be\label{eq:defnormal}
N_a = \p_a \Omega.
\ee
Its conformal frame norm\footnote{Once again, indices of unhatted quantities are raised with the conformal frame metric $g^{ab}$.}, $N^aN_a = \O^2$ becomes null at $\O=0$. 
 Note that $N_a$ is the inward-pointing normal, this will lead to a sign in the Stokes theorem. Some of the equations in the following will be simplified by the introduction of the normalized normal, which has unit modulus with respect to the conformal metric $g_{ab}$:
\begin{align}
n_a = \frac{1}{\Omega}N_a.
\end{align}

Working with the coordinate $\O$ and the metric $g_{ab}$, rather than $r$ and $\hat g_{ab}$, is useful because the components of $g_{ab}$ in the coordinates $(\O, u, x^A)$ are asymptotically finite. This framework will also automatically provide natural fall-offs for the fields and most importantly allow for a systematic analysis of the finiteness of asymptotic quantities.

Let $x^i = (u, x^A)$ be the coordinates on $\O = \const$ surfaces. The retarded Bondi coordinates $(\O,u,x^A)$ determine a coordinate projector $P_a^i$, which maps spacetime vectors to vectors on $\O = \const$ by dropping their $\po$ component. The exterior derivative can then be decomposed as
\begin{align}\label{eq:decomp_id}
\rd= \rd x^a \delta_a^b  \p_b =  \rd x^a P_a^i \p_i + N_a \rd x^a \po.
\end{align}
We will suppress the projector in the notation. For example $F^{ij} = P_a^i P_b^j g^{aa'} g^{bb'} F_{a'b'}$ is the projection of the field strength with raised spacetime indices, and not the pulled back field strength with indices raised by the inverse of the induced metric (which does not have a finite limit since $\scri$ is null in the conformal metric). Because $g^{u\O} =1$, $F^{ij}$ contains $F_\O{}^j$, and depends not only on the pullback of the gauge potential $A_i$ but also on $A_\O$ and the transverse derivatives of $A_i$. Note that since the projector is a coordinate projector, it commutes with coordinate derivatives, including $\po$.

The Bondi coordinates also determine a coordinate vector field $\po$, which is defined throughout the spacetime. This vector field will play a crucial role in the following. At $\scri$, $\po$ can be used to ``take orders in $1/r$'' of tensors, and in the next section, we will use this to introduce a Taylor expansion of the electromagnetic potential off of $\scri$ in powers of $\O$.
{We will focus on finite intervals in the retarded time $u$. This will allow us to discuss $u$-falloffs accurately}.  $I=I(\Omega=0)\subset\scri$ will denote the limit of the following hypersurfaces (with boundary)
\begin{align}
    I(\omega) = \{ (u, \O, x^A): \O = \omega, u_i \le u \le u_f \} 
    \quad\text{with}\quad
\p I(\omega) = S_i(\omega) \sqcup S_f(\omega),
\end{align}
where $S_i(\omega)$ ($S_f(\omega)$) is a codimension-2 sphere obtained as the cut of the hypersurface $\Omega=\omega$ at $u= u_i$, ($u=u_f$), respectively. At infinity, we will simply denote $\p I =S_i \sqcup S_f$.

%============================================================

\section{Equations of motion: asymptotic simplicity and the conformal current}
\label{sec:eomI}

Utilizing  Penrose's idea of asymptotic simplicity \cite{Penrose:1962ij, Penrose:1987uia, Penrose:1986ca}, we will assume that the components of the gauge field $A_a$ in the coordinates $(u, \O, x^A)$ have finite values at $\scri$ and admit an expansion in powers and log-powers of $\O$: 
\be
A_a = \sum_{k=0}^{D-4} \O^k A_{a(k)}  + \O^{D-5} P
\label{eq:expansion}
\qquad \text{where}\qquad
A_{a(k)}:= \frac{1}{k!} \po^k A_a{}_{|\O=0}.
\ee 
where the $P$ is a polyhomogeneous function of $\O$ (see appendix \ref{app:radial}).
We immediately get that all tensors and forms which are built from $A_a$ and $g_{ab}$, in particular $F_{ab}$ and $F^{ab}=g^{aa'}g^{bb'}F_{a'b'}$, are finite in the asymptotic limit and admit the same expansion. 
Stronger fall-offs on certain components of the fields will be automatically required by the equations of motion. Our falloffs admit radiative as well as Coulombic solutions.

It is important to appreciate that in general the analytic expansion breaks down and logarithms can appear. We prove finiteness of the renormalized symplectic potential even in the presence of logarithms and anomalies (see appendix \ref{app:radial}).  In particular, for our analysis, the above expansion does not need to be analytic to all orders, a requirement that would be physically extremely restrictive. 
{ It is enough to demand that $F_{ab}$ is $(D-5)$-polyhomogeneous (defined in apendix \ref{app:radial}) as the discussions of section \ref{sec:renormalize}, \ref{sec:EoMII} and appendix \ref{app:radial} will show.}
It is this differentiability requirement that will force us to restrict our analysis to even spacetime dimensions, for similar statements in gravity see also \cite{Hollands:2004ac, chrusciel2005boundary}.
To keep our notation as light as possible, finite (as opposed to infinite) differentiability will be left implicit in our formulas. 
Several works on relaxing analyticity in the gravitational case have appeared e.g. in \cite{winicour1985logarithmic, chrusciel1995gravitational,Friedrich2004,chrusciel2005boundary}, { while in 4D electromagnetism we refer to \cite{Herdegen:2016bio} and references therein}.

The Lagrangian of vacuum electrodynamics is

\begin{align}
\hat\bL &:= -\frac1{4} \hg \hat{F}^{ab}\hat{F}_{ab} =\O^{-(D-4)} \bL, \quad \text{ where }\quad
\bL  := -\frac14 \q F^{ab} F_{ab}.
\end{align}

where $A_a \equiv \hat A_a$, from which $\hat F_{ab} \equiv F_{ab}=\p_a A_b - \p_b A_a$ and $\hat F^{ab} = \hat g^{aa'}\hat g^{bb'}F_{a'b'} = \O^{4} F^{ab}$. $\bL$ has a finite limit onto $\scri$.

Varying $\hat\bL$ with respect to $A_a$ gives\footnote{We use boldface letters for spacetime densities and ``hats'' for unrescaled quantities referring to the physical spacetime. Hatted quantities can diverge in the limit $\O\to0$, while unhatted quantities are defined so that they will not. Geometrically, $\bM^a$ and $\hat \bth{}^a$ are codimension-1 forms, or densitized currents. }
\begin{equation}
\label{eq:deltaL}
\d \hat \bL ={} \hat \bM{}^a \d A_a + \p_a \hat\bth{}^a
\end{equation}
where $\hat\bth{}^a$ is the symplectic potential (SP) current density, which we will shortly come back to, and $\hat\bM^a$ are the equations of motion (EoM):
\begin{subequations}
\begin{align}
\hat\bM^a ={}& \p_b (\sqrt{\hat g} \hat F^{b a}) \\
={}& \O^{-(D-4)} \Big( \p_b(\q F^{ba}) - (D-4) \frac1\O N_b F^{ba} \Big).
\end{align}
\end{subequations}
Since $\p_b(\q F^{ba})$ is finite by assumption, in $D\neq4$, the dominant asymptotic order of the equations of motion comes from the second term. The dominant order is hence solved by requiring that $N_b F^{ba}$ is of order $\O$. We call these the \textit{asymptotic Maxwell conditions}
\be
(D-4)N_b F^{ba} \; \seq \; 0,
\label{eq:AMC}
\ee
and will require that they are implemented as a restriction on the field space itself. The asymptotic Maxwell conditions arise from demanding finiteness of the asymptotic fields $F^{ab}$, and will be crucial for the derivation of the ``soft'' part of the symplectic structure in section \ref{subsec:6D}.

The asymptotic Maxwell conditions allow us to define what we call the \textit{conformal current} as
\begin{align}\label{eq:em_conf_curr}
 \J^a := \frac1\O N_b F^{ba} \equiv F^{na},
\end{align}
where we recall $n_a = \O^{-1} N_a = \O^{-1} \p_a \O$.
The conformal current is defined throughout the spacetime,\footnote{By this, we mean at least in a collar neighborhood of $\scri$ itself.} not just on $\scri$ and, by the antisymmetry of $F$, it is tangential to the level surfaces of $\O$. By the asymptotic Maxwell conditions, it has a finite limit onto $\scri$ in $D\neq4$. This will be crucial in the following.

We can then rescale the EoM to remove negative powers of $\O$, obtaining
\begin{align}
 \bM^a  := \O^{(D-4)} \hat \bM^a ={}& \p_b(\q F^{ba}) - (D-4) \q\J^a.
 \label{eq:E}
\end{align}
In $D\neq4$, the vacuum EoM take the form of Maxwell equations in presence of an external source: the conformal current.\footnote{To avoid specifying further asymptotic properties, we neglect any matter contribution to the current. It seems however natural to require that the conformal current is a well defined quantity at $\scri$ even in presence of matter.} The origin of the conformal current is the fact that the EoM transform inhomogeneously under the conformal rescaling of the metric, or alternatively, that in the conformal frame Lagrangian $A$ is non-minimally coupled to a background scalar field $\O$. The normal component of the EoM reads
\begin{align}
  \bM^n = - \p_i (\q \J^i),
\end{align}
The conformal current is thus conserved on-shell.
This concludes the analysis of the EoM for now, we will come back to them in more explicit detail in section \ref{sec:EoMII}.

%=======================================================

\section{Renormalizing the symplectic potential}
\label{sec:renormalize}

The symplectic potential (SP) current density $\hat\bth{}^a$ determines the canonical structure of the theory. In the covariant Hamiltonian formalism \cite{Kijowski1976, crnkovic1987covariant, Gawedzki:1990jc,  Lee:1990nz}, which we use here, it is related to the Lagrangian through the equation \eqref{eq:deltaL}, which is usually taken to imply
\begin{subequations}
\begin{align}
 \hat \bth{}^a ={} \O^{-(D-4)} \bth{}^a 
\qquad\text{where}\qquad
 \bth{}^a :={} -\q F^{ab} \d A_b.
 \label{eq:def_theta}
\end{align}
\end{subequations}
We refer to $\hat \bth{}^a$ as the \textit{standard SP}. 
Its normal component, which determines the standard symplectic form on $I(\Omega)$, is\footnote{We suppress coordinate volume elements such as $\rd u \rd^{D-2} x^A$ if there is no risk of confusion.}
\begin{align}
    \hat \bth{}^\O = \O^{-(D-5)} \bth{}^n  \qquad \text{where} \qquad \bth{}^n = - \q \J^i \d A_i.
\label{eq:thetan}
\end{align}
The symplectic form on the $\O = \const$ surfaces is the integral of the (antisymmetrized) variation of its normal component.\footnote{In the language of differential forms, of its pullback on $I(\Omega)$.} Since $N_a$ is the inward facing normal, the integration comes with a sign. Hence, the contribution to the symplectic form from an interval $I(\Omega)$ is
\begin{align}
& \omega(\O) := \d \hat \Theta(\O) 
\end{align}
where
\begin{align}
& \hat \Theta(\O) := - \int_{I(\O)} \hat\bth{}^\O = - \O^{-(D-5)} \int_{I(\O)} \bth{}^n.
\end{align}
Notice that since $\O$ has a double role as ``canonical time'' for the radial evolution and as the conformal factor, the conformal current appears both as a source term in the EoM and in the SP as the momentum canonically conjugate to the tangential connection $A_i$ . $\bth{}^n$ has a finite, non-zero limit onto $\scri$.

In $D>4$, the SP on the level surfaces of $\O$ diverges as $\O^{-(D-5)}$ when approaching $\scri^+$. The divergence seems like bad news for the canonical theory, signifying potentially infinite Hamiltonians, infinite charge generators and ill-defined Poisson brackets. However, as is well known \cite{Jacobson:1993vj}, the SP is ambiguous. Firstly, adding a boundary term to the action adds a total variation to the SP (which does not change the symplectic form). Secondly, since $\hat\bth{}^a$ is defined only implicitly through \eqref{eq:deltaL}, it is ambiguous by the divergence of an antisymmetric tensor.
The ambiguities are
\begin{subequations}
\begin{align}
\label{eq:ambiguity}
\hat\bth{}^a & \mapsto \hat\bth{}^a + \p_b \hat{\boldsymbol \alpha}{}^{ab} + \d \hat{\boldsymbol \beta}{}^a
\end{align}
and thus
\begin{align}
\hat \bth{}^\O & \mapsto \hat \bth{}^\O + \p_i \hat \ba{}^{\O i} + \d \hat \bb{}^\O.\label{eq:ambigs_thetaO}
\end{align}
\end{subequations}
Here, $\hat\ba{}^{ab} = \hat\ba{}^{[ab]}$ is the \emph{corner counterterm}. It is a codimension-$2$ density and it modifies the canonical expression of the boundary charges. $\hat\bb{}^a$ is a change of polarization coming from a choice of  \emph{boundary action},  $\hat \bL \mapsto\hat \bL + \p_a \hat \bb{}^a$. 
The normal component $\hat \bth{}^\O$ is only defined up to a total variation and a corner term.  
The modification of $\hat \bth{}^\O$ by a total derivative $\p_i \hat\ba{}^{\O i}$ represents the addition of a corner term to the symplectic potential. 
The corner term $\hat\ba^{\Omega i}$ appears in the redefinition of the asymptotic charges.

We can now phrase the main idea behind our construction.
In order to have a well defined action on an asymptotically simple spacetime and a finite symplectic structure at $\scri$, what really matters physically  is that it is possible to reabsorb all the divergences of $\hat\bth{}^a$  into a divergent boundary action and  divergent corner terms.
We now show that this is exactly the case.  
As already hinted, this procedure simultaneously renormalizes the action and all the Noether charges. Therefore, it generalizes the holographic renormalization procedure of AdS/CFT \cite{de2001holographic,  papadimitriou2010holographic}  to asymptotically flat spacetimes.
The critical difference here is the necessity to renormalize not only the action but also renormalize the soft charges that generate the asymptotic symmetry algebra. 
Renormalization of both the soft charges and the action amounts to the definition of a finite asymptotic symplectic potential. 
We now turn to this task.

 Splitting the divergence in the defining relation \eqref{eq:deltaL} for the SP, $\d \hat \bL = \hat \bM{}^a \d A_a + \p_a \hat \bth{}^a$, into a divergence on the $\O = \const$ surfaces and a transverse derivative by using the decomposition \eqref{eq:decomp_id} of the identity, one obtains
\begin{align}
  \d\hat\bL =   \d(\O^{-(D-4)} \bL) = \O^{-(D-4)} \bM^a \d A_a + \O^{-(D-4)} \p_i \bth{}^i + \po (\O^{-(D-5)} \bth{}^n).
\end{align}
Rearranging the terms and extracting the factor in $\O$, one obtains
the asymptotic renormalization equation: 
\begin{align}\label{eq:re_theta}%\boxed{
\big[ (D-5) - \O \po \big] \bth{}^n = \p_i \bth{}^i - \d \bL + \bM^a \d A_a.
%}
\end{align}
This equation for the normal SP is  the key to our main result. Crucially, modulo EoM, the RHS contains only a total derivative and a total variation which are part of the ambiguity in $\hat\bth{}^\O$. We call equations involving the operators $(n - \O\po)$ \textit{radial equations}. Their general properties are discussed in appendix \ref{app:radial}. { Notice that demanding the gauge potential to be $(D-4)$-polyhomogenous as above, ensures the field strength and thus the right hand side of this equation to be $(D-5)$-polyhomogeneous, as required in the theorems of the appendix.}

The radial equation \eqref{eq:re_theta} implies that $\hat \bth{}^\O$ can be made finite on-shell by subtracting counterterms which fall under the ambiguities \eqref{eq:ambigs_thetaO}. As a first way to see it, note that at each order of a Laurent series for $\hat \bth{}^\O$ 
%\begin{subequations}
\begin{align}
  \hat \bth{}^\O  = \O^{-(D-5)} \bth{}^n_{(0)} + \O^{-(D-6)} \bth{}^n_{(1)} + ... + \O^{-1} \bth{}^n_{(D-6)} + \bth{}^n_{(D-5)} + \O \bth{}^n_{(D-4)} + ...%\\
\end{align}
the radial equation reads
\begin{align}
  (D-5-k) \bth{}^n_{(k)} \; \oeq \; \p_i  \bth{}^i_{(k)} - \d \bL_{(k)},
\end{align}
%\end{subequations}
where $\oeq$ denotes on-shell equality. 
The orders $k < D-5$ of $\bth{}^n$, which are the ones that come with divergent prefactors in $\hat \bth{}^\O$, are  fixed on-shell by the radial equation to be total derivatives plus total variations, while $\bth{}^n_{(D-5)}$, which gives the finite order of $\hat\bth{}^\O$, is not determined. The remaining terms do not contribute in the asymptotic limit $\O\to0$. Thus, it  is clear that the divergences in $\hat \bth{}^\O$ can be removed order by order in the Laurent series. 

Rather than working order by order, we perform the counterterm subtraction at finite distance and take the limit in the end. 
In this way we obtain the asymptotic SP as the finite limit of a renormalized SP.

Applying the results of appendix \ref{app:radial} to equation \eqref{eq:re_theta}, one obtains the renormalized normal-component to the hypersurface $\Omega=\const \geq0$ in the form\footnote{Recall that in our notation: $X_{(k)}:=\tfrac1{k!} \po^k X_{|\O=0}$. \label{ftntnotation}}
\begin{align}
\hat\bth{}^\O_R & = \O^{-(D-5)}\bth^n_R \notag\\
&= \hat\bth{}^\O - \O^{-(D-5)}\sum_{p=1}^{D-5} \Big[ \tfrac{(D-5-p)!}{(D-5)!}\O^{p-1}\po^{p-1}\big( \p_i\bth^i - \delta \bL\big)\Big] + \ln \O \;  (\p_i\bth^i_{(D-5)} - \delta \bL_{(D-5)}).
\label{eq_thetanR}
\end{align}
{ The main claim that follows from the analysis done in the appendix (see eq. \eqref{Xexp}) is that  
$\hat\bth^\O_R$ admits a finite limit at $\Omega=0$ as long as $\bth^i$ and $\bL$ are
$(D-5)$-polyhomogeneous.
 This means that we assume that the  fields, and hence the Lagrangian and tangent symplectic potential, have an expansion of the form
$L= \sum_{k=0}^{D-5} L_{(k)} \O^k +
\O^{D-4} P(\O)$, where $P$ is polyhomogeneous in $\O$. }
This readily translates into the following renormalization prescription for the full SP:
\begin{align}
    \hat \bth{}^a_R :={}& \hat \bth{}^a + \p_b \hat \ba{}^{ab} + \d \hat \bb{}^a,
\end{align}
with corner counterterms
\begin{subequations}
\begin{align}
\hat\ba{}^{\Omega i} &= - \hat\ba{}^{i\O} = - \O^{-(D-5)}\sum_{p=1}^{D-5} \Big[  \tfrac{(D-5-p)!}{(D-5)!}\O^{p-1}\po^{p-1}\bth^i\Big] +   \ln \O \;  \bth^i_{(D-5)} \\
\hat\ba{}^{ij}&=0
\end{align}
\end{subequations}
and boundary action
\begin{subequations}
\begin{align}
\hat\bb{}^{\Omega } & = \O^{-(D-5)}\sum_{p=1}^{D-5} \Big[  \tfrac{(D-5-p)!}{(D-5)!}\O^{p-1}\po^{p-1}\bL\Big] -  \ln \O \;  \bL_{(D-5)} \\
\hat\bb{}^{i}&=0
\end{align}
\label{countermermbeta}
\end{subequations}
The choice $\hat\ba{}^{ij}=0=\hat\bb{}^i$ is not unique, and can be modified without interfering with the renormalization of the SP on $\scri$. However, notice that even with this choice, $\hat\bth{}^i_R$ is nonetheless renormalized by  $\po \hat\ba{}^{i\O}$.

It is crucial to keep present the following two facts about the above renormalization procedure. 
On the one hand, the algebraic part of the counterterms is given by forms on spacetime and not only on $\scri$, i.e. by expressions local in all coordinates including in the coordinate $\O$; in particular, these counterterms are {\it not} truncated Laurent series in $\O$.
On the other hand, the coefficients of the logarithmic terms are given by derivatives $\po^{(D-5)} \bth^i$ and $\po^{(D-5)}\bL$ {\it evaluated at} $\O=0$.\footnote{See previous footnote, fnt \ref{ftntnotation}.}
If we assume analyticity of the field expansion, the logarithmic terms have to vanish and the expansion is purely algebraic. It is convenient to combine the logarithmic term into what we call the {\it SP} ({\it logarithmic}) {\it anomaly}:
\begin{align}
\mathcal C := \p_i\bth^i_{(D-5)} + \delta \bL_{(D-5)}.
\label{eq:SPanomaly}
\end{align}
We call the terms  $\p_i\bth^i_{(D-5)}$ and $\bL_{(D-5)}$  the charge and action anomaly respectively.

A more explicit expression for the SP on $I\subset\scri$,\footnote{The minus sign is due to the ingoing direction of the normal to $\scri$.}
\be
 \hat\Theta{}^{R} := - \lim_{\O\to0} \int_{I(\O)} \hat\bth{}^\O_R (\O) ,
\ee 
can be found using the second result of appendix \ref{app:radial}, equation \eqref{Xexpd}, which gives
\be
\hat\bth{}^\O_R (\O) = \O^{-(D-5)} \bth^n_R(\Omega) =  \tfrac{1}{(D-5)!} \po^{D-5} \bth^n + \ln \O \;\mathcal C .
\ee 
Distributing the radial derivative on $\bth^n$ as given in \eqref{eq:thetan} and taking the $\O\to0$ limit gives the following expression for the renormalized SP on a region $I\subset\scri$:
\begin{align}
\label{SPsum}
    \hat\Theta{}^{R} =  
    \sum_{k = 0}^{D-5} \Theta^{R}_{(k)} + H_{D-5}  \int _I \mathcal C,
  \qquad
        \Theta^{R}_{(k)} : = \int_I \q \;\J^i_{(D-5-k)} \d A_{i (k)}.
\end{align}
The sum involves $D-4$ terms associated with different ``layers'' of the conformal current, from $\J_{(D-5)}$ to $\J_{(0)}$.
These layers are dynamically interdependent, a fact that we will analyze in detail in section \ref{sec:EoMII}. 
In the following we will consider only analytic solutions (up to an appropriate order)  in which $\mathcal C=0$. Even if it were not vanishing, though, the SP anomaly could still be reabsorbed in the ambiguity of $\hat\bth{}^\O_R$.
$H_{n} = \sum_{p=1}^n p^{-1}$ is the $n$-th harmonic number.

To conclude this section, let us stress the role played by the counterterms $\hat\ba{}^{ab}$ and $\hat\bb{}^a$. Whereas the physical interpretation of $\hat\bb^a$ is clear---it is meant to renormalize the action---the interpretation of $\hat\ba^{ab}$ may seem more mysterious. However, its role is physical and is meant to renormalize the symmetry generators, i.e. the Noether charge, associated to the (asymptotic) gauge symmetries.
We now turn to their analysis.

%=======================================================

\section{Generators} 
\label{sec:generators}

The on-shell generators of gauge transformations are a crucial ingredient for the interpretation of asymptotic symmetries. We present  part of their evaluation in this section. As stated above, the data in the  SP are not all independent---however, this is no impediment to the calculation below: if one is interested just in the charges, resolving the dependencies can be delayed until after an expression for the charges has been obtained, streamlining the computation.

To get the generators for the renormalized symplectic form, one could start from the standard generators associated to the standard symplectic form, and calculate how they change due to the corner counterterms.\footnote{{ As opposed to the corner counterterms $\hat\ba$,} the boundary action $\hat\bb$ is built out of $\bL$ and is therefore manifestly gauge invariant. As the following derivation shows, this means that it does not contribute to the { renormalization of the Hamiltonian generators -- even though it does contribute to the renormalization of the symplectic potential.}} Alternatively, one can calculate the generators from the renormalized SP directly. We will take the second route. We perform the calculation directly at $\scri$, but since the renormalized symplectic form is known at finite distance one can in principle do the same computation there.

The asymptotic renormalized symplectic form is
\begin{align}
    \hat\omega^R := \delta \hat\Theta^R ={}& \int_I \sum_{k=0}^{D-5} \d \J_{(k)}^i \curlywedge \d A_{i(D-5-k)},
\end{align}
where $\curlywedge$ denotes antisymmetrization of the $\delta$'s.
The generators of gauge transformations $\hat H_\alpha^R$ are related to the symplectic form as
\begin{align}
    \d \hat H^R_{\epsilon} = - I_\epsilon \hat\omega_R
\end{align}
where $I_\epsilon$ is the action of a gauge transformation, i.e., $I_\epsilon \hat\omega_R (\d A_a, \d A_a) = \hat\omega_R (\p_a \epsilon, \d A_a)$. The action of a gauge transformation $\epsilon =  \epsilon_{(0)} + \O \epsilon_{(1)} + ...$ on the variables is
\begin{align}
    I_\epsilon \d A_{i(k)} = \p_i \epsilon_{(k)},
    \qquad
    I_\epsilon \d A_{\O(k)}=(k+1)\epsilon_{(k+1)},
    \qquad 
    I_\epsilon \d \J^i_{(k)} = 0.
\end{align}
Using the conservation of the conformal current $\p_i (\q \J_{(k)}^i) = 0$, we obtain for the asymptotic renormalized on-shell generators
%\begin{subequations}
\begin{align}
    \hat H^R_{\epsilon} ={}& \big[ \hat Q^R_\epsilon]^f_i
\qquad\text{where}\qquad
 \hat Q^R_\epsilon ={} \sum_{k=0}^{D-5} \oint_S  \q \J^u_{(D-5-k)} \epsilon_{(k)}
\label{eq:renorm_hamiltonian}
\end{align}
%\end{subequations}
and $[X]^f_i := X(u_f) - X(u_i)$.
This expression is manifestly finite, and should be contrasted with the generators obtained from the standard symplectic form $\hat\omega = \O^{-(D-5)} \int_{I(\O)} \q \d \J^i \curlywedge \d A_i$, which read $ \hat H_{\epsilon} = [\hat Q_\epsilon]^f_i$ with
\begin{align}
    \hat Q_{\epsilon}(\O) = \O^{-(D-5)} \oint_{S(\O)} \q \J^u  \epsilon 
\end{align}
and diverge in the asymptotic limit (unless one puts strong restrictions on the space of asymptotic data). 

Observe that just as the renormalized SP coincides asymptotically with the finite part of the Laurent series of the standard SP, the renormalized generators are the finite part of the standard generators. 

The ``layering'' structure also transfers from the SP to the charges: there are not one, but $(D-4)$ ``sphere-worth'' of non-zero charges, which depend on the extension of $\epsilon$ off of $\scri$. The extension dependence of charges has been noted for the gravitational case already in \cite{Geroch:1981ut}.
We will revisit the layering and the extension ambiguity of the charges in section \ref{sec:softvsgauge}.

%============================================================

\section{Asymptotic equations of motion}
\label{sec:EoMII}

In this section, we give the complete set of relations between the quantities entering the renormalized SP. Specifically, we will identify the free data needed to solve the EoM asymptotically. Computations are performed in general $D\ge 6$ (even).

The first step is to split the EoM into their radial, retarded-time, and sphere components, and hence to develop them in orders of $\O$. We will write the equations in ``radial-time'' gauge\footnote{Since $\scri$ is null and transverse to $\po$, the radial gauge $A_\O\equiv0$ shares there various features with the usual time gauge $A_t\equiv 0$ fixed at a standard Cauchy surface $\Sigma_{t=\const}$.}
\be
\qquad A_\O \equiv 0,
\label{eq:gf}
\ee
and will comment in section \ref{sec:softvsgauge} on the status of that condition and how to lift it.

Consider first the conformal current $\J^a = \O^{-1} N_b F^{ba}$.
We write the definitions of $\J^u$ as a radial evolution equation for $\A_u$ and the definition of $\J^A$ as a retarded time evolution equation for $A_A$:
\begin{subequations}
\begin{align}
    \po A_u &= - \O\J^u \\
    \p_u A_A &= \p_A A_u + \O\J_A - \O^2\po A_A.
\end{align}
\label{eq:Ngfixed}
\end{subequations}
The EoM, $\bM^a ={} \p_b(\q F^{ba}) - (D-4)\q \J^a$, can be decomposed as:
\begin{subequations}
\begin{align}
   \bM^n ={}& - \p_u(\q \J^{u}) - \p_A(\q \J^{A}),\\
    \bM^u ={}& - \Big[ (D-5) - \O\po \Big](\q \J^{u}) - \p_A\po(\q A^{A}), \\
    \bM^A ={}& - \Big[ (D-5) - \O\po \Big](\q \J^{A}) + \p_u\po(\q A^{A}) + \p_B(\q F^{BA}) \label{eq:bMA_1} \\
    ={}& - \Big[ (D-6) - 2\O\po\Big](\q \J^{A}) + \p_B(\q F^{B A}) - \q(1+\O\po)(\O\po A^A) \nonumber \\
    & - \O \q \p^A \J^u,
\end{align}
\end{subequations}
where in the last line we have rewritten $\bM^A$ as a purely radial evolution equation, by means of \eqref{eq:Ngfixed}.
Notice the factor of 2 which appeared in the radial derivative operator as a consequence of this manipulation.
We will come back to it shortly.

We now develop the equations in orders of $\O$. First, consider the normal component of the EoM,
\begin{align} \label{eq:bMOk}
    \bM^n_{(k)} ={}& - \p_u (\q \J^u_{(k)}) - \p_A (\q \J^A_{(k)}).
\end{align}
Note that the identity $\p_a \p_b (\hg \hat F^{ab}) = 0$ can be written as
\begin{align}
 \Big[ (D-5) - \O\po \Big]\bM^n = \p_u\bM^u + \p_A\bM^A
\end{align}
Asymptotically, this implies that the only independent information contained in  $\bM^n$ lies in its $k=(D-5)$ order. The rest of its orders automatically vanish once the tangential EoM are solved, and do not need to be considered separately. Thus we define
\begin{subequations} \label{eq:complete_as_equations}
\be \label{eq:gaussk}
    {\bf G}:= \bM^n_{(D-5)} = -\p_u(\q \J^u_{(D-5)}) - \p_A(\q \J^A_{(D-5)}).
\ee
As it will become clear shortly, this is the Gauss law on the $\O = \const$ slices. 
The orders of the remainder EoM and the definitions of the conformal current are
\begin{align}
    A_{u(k+1)} &= - \tfrac{1}{(k+1)}\J^u_{(k-1)} \label{eq:Auk+1}\\
    \p_u A_{A (k)} &= \p_A A_{u (k)} + \J_{A (k-1)} - (k-1) A_{A(k-1)} \label{eq:puAAk}\\
    \bM^u_{(k)} ={}& - (D-5 - k)(\q \J^{u}_{(k)}) - (k+1)\p_A(\q A^{A}_{(k+1)}) \label{eq:Euk}\\
    \bM^A_{(k)} ={}& - (D-6 - 2k)(\q \J^{A}_{(k)}) + \p_B(\q F^{B A}_{(k)}) - \q k(1+k) A^A_{(k)} - \q \p^A \J^u_{(k-1)}.\label{eq:EAk}
\end{align}
\end{subequations}
These equations hold for $k \ge 0$ if we set negative orders of $\J$ and $A$ to zero by convention. 
The equations \eqref{eq:complete_as_equations} are the complete set of asymptotic EoM.

These equations contain the asymptotic Maxwell conditions $N_a F^{ab}\;\seq\; 0$, which are explicitly given by 
\begin{align}\label{AsymE}
    A_{u(1)} = 0, \qquad \J^u_{(-1)}=0,\qquad \p_u A_{A(0)} = \p_A A_{u(0)}.
\end{align} 
The last equation can be conveniently solved by introducing a Hodge decomposition of
\be
A_{A(0)}=\epsilon_A{}^{BC\cdots}\p_B\mu_{C\cdots}+\p_A\varphi =:\alpha_{A(0)} + \p_A \varphi .\label{Fn12}
\ee
Then, equation \eqref{AsymE} says that the purely magnetic part $\alpha_{A(0)}$ must be $u$-independent and that the purely electric part $\varphi$ is related to $A_{u(0)}$ by
\be
A_{u(0)} =  \p_u\varphi. \label{softpot}
\ee
We call $\varphi$ the \emph{soft potential}.\footnote{Notice that $\varphi$ in \eqref{softpot} is not fully determined by the Hodge decomposition of $A_{A(0)}$, but only up to a time-dependent sphere-constant term. We will see that $\varphi$ is in an appropriate sense canonically conjugated to the local electric flux. Thus, since in absence of charged matter the total flux vanishes, this sphere-constant term does not play much of a role, see section \ref{sec:sp7}.}

We are now going to analyze the asymptotic EoM to identify the asymptotically free data. 
As before, we focus on a finite region $I \subset \scri$, with $u_i \le u \le u_f$. 
The boundary of $I$ is the union of two \emph{corner} spheres, 
denoted $\p I = S_i \sqcup S_f$, where $S_{i}$ ($S_f$) is the cut of $\scri $ at $u=u_i$ ($u=u_f$, respectively).

We view the conformal current $\J^i$ as an a priori independent variable from the gauge field, such that the definition of $\J^i$ in terms of components of the gauge field has the same status as the EoM. While this is not strictly necessary for electromagnetism, it can potentially clear up the analysis of the EoM in the gravitational case. The key to identifying the free data is that the factor $(D-5-k)$ in \eqref{eq:Euk} becomes zero for $k=D-5$, and the factor $(D-6-2k)$ in \eqref{eq:Euk} becomes zero for $k = \frac{D-6}{2}$. For later convenience we introduce the new symbol
\be
\ell :=\frac{D-6}{2}.
\ee
Note the obvious relations $D=6 + 2\ell$ and $D-5=2\ell+1$.

We will first state how to solve the EoM iteratively and what the free data are at the ``generic'' orders $k \notin \{0,\ell, 2\ell +1 \}$, and return to those three orders below.
It is also useful to define $ \alpha_{A (k)} := A_{A(k)}(u_i)$
which is a corner variable evaluating the value of $A_A$ at the initial slice. The value of $A_A$ on a arbitrary time slice can then be obtained as
\be\label{Auint}
A_{A (k)}(u)=\alpha_{A\, (k)} +\int_{u_i}^u \p_u A_{A (k)}(u') \rd u'. 
\ee

We are now in a position to show that the free canonical data on $I$ 
is given by 
\be
\left\{ \varphi,\, \J^A_{(\ell)} \right\} \,\, \mathrm{ on}\,\, \scri, \qquad \left\{ \J^u_{(2\ell+1)},\, \alpha_{A(0)},\cdots ,\alpha_{A(2\ell+1)} \right\} 
\,\, \mathrm{ on}\,\, S_i,
\ee
and we prove this by recurrence.
We start the recurrence by assuming that we are given the variables 
$A_{u(0)}$ and $\alpha_{A(0)}$. 
Equation \eqref{AsymE}, determines $\p_uA_{A(0)}$ hence 
$A_{A\,(0)}$.
To continue the recurrence it is convenient to lay out the equation of motions as follows\footnote{$D_A$ is the covariant derivative on the sphere $S$, so that e.g. $\p_A(\sqrt{q} v^A) = \sqrt{q}D_Av^A$.}
\begin{subequations}\label{simpleeq}
\bea
  (D-4 - k) \J^{u}_{(k-1)} &=& - kD_A A^{A}_{(k)},
     \label{eq:Euk2}\\
   (D-6 - 2k) \J^{A}_{(k)} &=& D_BF^{B A}_{(k)} - [\p^A \J^u_{(k-1)} +k(k+1) A^A_{(k)}]   ,\label{eq:EAk2}\\
          \p_u A_{(k+1)}^A &=&  \J^A_{(k)} -\tfrac1{k+1} [\p^A \J^u_{(k-1)} +k(k+1) A^A_{(k)}] . \label{eq:puAAk2}
\eea
\end{subequations}

We now assume that $A^A_{(k)}$ is known on $I$.
The first equation defines $\J^{u}_{(k-1)}$ from $A^{A}_{(k)}$,
as long as $k\neq (D-4)$, 
the second defines $\J^{A}_{(k)}$ from $(A^{A}_{(k)},\J^{u}_{(k-1)})$,
as long as  $ k \neq \ell$ and the third determines $\p_u A^A_{(k+1)} $ from $(A^{A}_{(k)},\J^{u}_{(k-1)},\J^{A}_{(k)} )$.  This in turns determines $A^A_{(k+1)}$ from $\alpha^{A}_{(k+1)}$ and \eqref{Auint} and we can start a new cycle of recurrence.\footnote{ The knowledge of $A_{u(k+1)}$ for $k>1$ is not explicitly required, one just deduce its value from $A_{u(k)} = -\J_{u (k-1)}/ (k+1)$}

This establishes that the free data is $\{\varphi, \J^A_{(\ell)}, \J^u_{(2\ell+1)}\}$ on $I$ and $\{\alpha^A_{(k)}\}$ on $S$.
One can then use the Gauss law to deduce the value of 
$\p_u \J^u_{ (2\ell+1)}$. This effectively reduces the free part of $\J^u_{ (2\ell+1)}$ to its initial value on $S_i$.

We conclude this section with a remark.
So far, whenever only the retarded-time derivative of a quantity $A$ was determined by the equations of motion, we have introduced an integration constant $\alpha$ associated to the initial value of $A$, i.e. $\alpha=A(u_i)$. Of course, this association is somewhat arbitrary: provided one had accordingly changed the integration kernel of $\p_uA$, one might have chosen $\alpha$ to be e.g. the final value of $A$, or the  ``zero-mode'' component , $\langle A \rangle := \int^{u_f}_{u_i}  A(u) \rd u $ (see appendix \ref{app:Green} for the integration kernel associated to this choice of $\alpha$). This freedom turns out to be useful when inspecting the symplectic structure of the theory.

%============================================================

\subsection{News, charge aspects, and radiative modes\label{sec:newsdef}}

As we have seen there are two currents that are exceptional in the sense that they are not determined recursively by the rest of the data.
The first exception appears at  order $k=\ell$, with $\ell:= \tfrac{D-6}{2}$: the variable $\J^A_{(\ell)}$ is not determined by \eqref{eq:EAk2}, contrarily to its other orders which are algebraically determined. It is free data on all of $\scri$. We call it the {\it Maxwell news}:
\begin{align}
    \N^A := \J^A_{(\ell)},
\end{align}
for its role in the asymptotic EoM is analogous to the Bondi news in 4D General Relativity. It is the free radiative data. Let us further introduce, the \emph{radiative modes}
\begin{align}
    \mathcal{A}_A := A_{A(\ell+1)}.
\end{align}
Using \eqref{eq:puAAk2}, $\mathcal A_A$ is determined by $\N^A$, up to an integration constant
\be\label{News1}
\N^A = \p_u \A^A + \ell  \left(  A^A_{(\ell)}-\frac{\p^A (D_B A^B_{(\ell)})}{(\ell+1)(\ell+2)}  \right).
\ee

In odd spacetime dimensions, all orders of $\J^A_{(k)}$ are algebraically determined by \eqref{eq:EAk}. We thus see from an asymptotic perspective that in odd spacetime dimensions, solutions which are ``smooth'' around $\scri$ do not have free radiative data. This is why we restrict our analysis to even dimensions. A similar statement has been made, albeit from a different perspective, for gravity e.g. in \cite{Hollands:2003ie,chrusciel2005boundary}.

For the last exception, consider the order $k = 2\ell+1 =D-5$, where the factor in \eqref{eq:Euk} vanishes. $\J^u_{(2\ell+1)}$ is hence not determined by \eqref{eq:Euk}, unlike the other orders of $\J^u_{(k)}$ which are algebraically determined. The retarded time evolution is, however, determined by the Gauss law \eqref{eq:gaussk}. We hence call 
\begin{align}
    \sigma := \J^u_{(2\ell+1)}
\end{align}
the \emph{charge aspect}, for its role is analogous to the (Bondi) mass aspect in general relativity. Note also that asymptotic Coulombic fields, such as the spherically symmetric Coulombic field of a finite point charge in the interior of spacetime, fall off such that they contribute to $\sigma$, but not to the orders $\J^u_{(k < D-5)}$. 

The charge aspect conservation is controlled by the Gauss law \eqref{eq:gaussk},
\be
\p_u \sigma + D_A \J^A_{(D-5)}=0.
\ee
This can be more explicitly expressed by using \eqref{simpleeq}, and taking the divergence of \eqref{eq:EAk2}, as
\be\label{Gauss2}
\p_u \sigma = \frac{D-5}{D-4} \Big(D^A D_A - (D-4)\Big) (D_B A^B_{(D-5)}). 
\ee

In $D=6$, this readily gives a relation between the conservation of the charge aspect and the radiative modes: 
\be
\p_u \sigma = \tfrac12 \left(D^A D_A - 2\right) (D_B \mathcal A^B) \qquad (D=6). 
\ee
However, in general $A_{A(D-5)}$ does not correspond to the radiative modes, and one might wonder whether a relation analogous to this one still holds in general dimensions (this relation is crucial for the derivation of the soft theorems, see \cite{Kapec:2014zla}). 
Indeed, a similar relation exists, but it rather expresses $\p_u^{\ell+1} \sigma$ in terms of spatial derivatives\footnote{By ``spatial derivative'' we mean derivative along the sphere, i.e. $D_A$.} of $\mathcal A_A$. 
This relation can be found by taking the divergences of equations \eqref{eq:EAk2} and \eqref{eq:puAAk2}. 
To see this, it is convenient to rewrite equations \eqref{simpleeq} for $k\neq2\ell+2$ as
\begin{subequations}\label{simpleeq2}
\bea
 \J^{u}_{(k-1)} &=& - \tfrac{k}{(2\ell+2-k)} D_A A^{A}_{(k)},
     \label{eq:Euk3}\\
  2(\ell-k) \J^{A}_{(k)} 
  &=& D_BF^{B A}_{(k)} +   \tfrac{k}{(2\ell+2-k)} \big[    D^A D_B-(k+1)(2\ell+2-k) \delta^A_B \big] A^B_{(k)} ,
  \label{eq:EAk3}\\
    \p_u A_{(k+1)}^A &=&  \J^A_{(k)}  
          +\tfrac{k}{(k+1)(2\ell+2-k)} \big[ D^A D_B-(k+1)(2\ell+2-k) \delta^A_B \big]  A^B_{(k)}. \label{eq:puAAk3}
\eea
\end{subequations}
where
\be
c^\ell_k :=(k+1)(2\ell+2-k)
\ee
is a symmetric coefficient under the exchange $k \leftrightarrow 2\ell+1-k$.

Thus, the divergences of \eqref{eq:EAk2} and \eqref{eq:puAAk2} readily give a recursion relation\footnote{With similar methods, a recursion relation can be found for $F^{AB}_{(k)}$ by taking the antisymmetric derivative of equations  \eqref{eq:EAk2} and \eqref{eq:puAAk2}, instead of their divergences.
} 
for $D_A A^A_{(k)}$: 
\begin{subequations}
\begin{align}
D_A \J^A_{(k)} ={}& \tfrac{k}{2(\ell-k)(2\ell+2-k)}\Dk{k} (D_A A^A_{(k)})\\
\p_u(D_A A^A_{(k+1)} ) ={}& \tfrac{k (2\ell + 1 -k)}{ 2 (\ell - k)(k+1)(2\ell + 2 -k) } \Delta_k^\ell (D_A A^A_{(k)})
 \end{align}
\end{subequations}
where we introduced the elliptic negative-definite differential operator
\be
\Dk{k} := \Big( D_A D^A - (k+1)(2\ell +2-k) \Big).
\ee
Using the above recursion relation, we find
\bea
\p_u^{\ell+1}\sigma &=& -\p_u^\ell( D_A\J^A_{(2\ell+1)}) = \tfrac{2\ell+1}{2(\ell +1)}
\Dk{2\ell+1} \p_u^\ell (D_AA^A_{(2\ell+1)})\cr
&=&\cdots= \frac{(-1)^{\ell}}{2^{(\ell+1)}} \frac{1}{ (\ell+1) !} \Big(
\Dk{2\ell+1}
\Dk{2\ell}
\cdots 
\Dk{\ell+1}\Big)(D_A \A^A).
\eea\label{eq:gaussell+1}
Thus, we see that in dimensions $D>6$ (even), i.e. $\ell>0$, the radiative potential only controls the higher time derivative $\p^{\ell+1}\sigma$ of the charge aspect.
It is for this reason that we need the intermediate potentials $\{A_{A(\ell+1+p)}\}_p$, as these control the lower derivatives
$\p^{\ell+1-p}\sigma $, for $p=1,\cdots, \ell$.

%=================================================================

\subsection{Anomalies}\label{anomaly2}

Even though the equations $\bM^A_{(\frac{D-6}{2})}\equiv\bM^A_{(\ell)}=0$ and $\bM^u_{(D-5)}\equiv\bM^u_{(2\ell+1)}=0$ do not determine the Maxwell news and the charge aspect, they of course still hold true.
Similarly to the SP radial equation \eqref{eq:re_theta} which is controlled by the logarithmic anomaly $\mathcal C$ of equation \eqref{eq:SPanomaly}, the $\bM^A$ and $\bM^u$ equations for $\J^A$ and $\J^u$ are also controlled by their own logarithmic anomalies. We will call them the vector and scalar anomaly respectively.
The vanishing of the vector and scalar anomalies corresponds to the smoothness of the conformal currents $\J^A$ and $\J^u$ (at least up to the orders $k=\ell$ and $k=2\ell+1$, respectively). We emphasize that this assumption of vanishing anomalies is not necessary for the finiteness of the renormalized symplectic potential.  
We assume their vanishing, as done in most of the literature, to simplify the expressions of the fields in terms of the asymptotic free data.

The vanishing of the scalar anomaly gives 
\begin{align}\label{eq:ana2}
    D_A A^A_{(2\ell+2)}=0.
\end{align}
This condition is {\it not} a restriction on the asymptotic data, since only the values of the potential $\{A^A_{(0)},\cdots, A^A_{(2\ell+1)}\}$ enters the definition of the SP.

The vector anomaly is controlled by the order $k=\ell:=\frac{D-6}{2}$.
Its vanishing hence restricts $A^A_{(\ell)}$. 
Using that  $\J^u_{(\ell-1)}=- \frac{\ell}{\ell+2} D_B A^B_{(\ell)}$,  one gets 
 \be \label{eq:vectoranomaly}
 D_B F^{AB}_{(\ell)}   + \ell \left[
 (\ell+1) A^A_{(\ell)} - \tfrac{1}{\ell+2}  \p^A (D_B A^B_{(\ell)})
\right] = 0.
 \ee
In $D\neq 6$, taking the divergence  of this equation we get that $[D_A D^A - (\ell+1)(\ell+2) ] (D_B A^B_{(\ell)}) = 0$. 
The Laplacian on the sphere has negative eigenvalues so this equation implies $ D_B A^B_{(\ell)}=0$ and  $ D_B  F^{AB}_{(\ell)}=\ell  (\ell+1) A^A_{(\ell)}$.
Using the relation \eqref{eq:Euk2}, we can translate the first condition into a  restriction on  $\J^u_{(\ell-1)}$. Hence, the vanishing of the vector anomaly implies 
 \begin{align}\label{vecanomaly}
 \J^{u}_{(\ell -1)} ={} 0 \quad\text{and}\quad 
 D_B  F^{AB}_{(\ell)}={}\ell  (\ell+1) A^A_{(\ell)}.
\end{align}

In dimension $D=6$, $\ell=0$ and the above manipulations fail. However, the two equations (\ref{vecanomaly}) stay true: the first one degenerates with the asymptotic Maxwell condition \eqref{AsymE}, while the second one simply means that $D_B F^{AB}_{(\ell=0)}=0$, compatibly with \eqref{eq:vectoranomaly}. By Hodge theorem, this equation is equivalent to\footnote{In \cite{Kapec:2014zla}, the same condition in arbitrary dimension $D>6$ is derived from a finite energy argument. Since we have here renormalized the symplectic form, the generators of time translations, whose on-shell value is energy, are likewise renormalized and their argument cannot be directly applied.}
\be
F^{AB}_{(\ell=0)}=0 \qquad (D=6),\label{F6}
\ee
an equation that holds only in $D=6$.
Furthermore, in all dimensions, the vanishing of the vector anomaly simplifies the expression \eqref{News1} for the news tensor, giving
\be\label{News2}
\N^A = \p_u \A^A + \ell    A^A_{(\ell)} 
= \p_u \A^A +\tfrac1{(\ell+1)}D_B F^{AB}_{(\ell)}.
\ee 

We leave the full analysis of these anomalous relations in the general context where we do not impose analyticity to future work.

%===================================================================

\section{Renormalized symplectic potential}
\label{sec:sp7}

With these results, we can now analyze  the renormalized asymptotic SP in the case where $D=6+2\ell\geq 6$ and even.\footnote{As we have seen, in odd dimension the analyticity conditions implies that there are no compatible radiative data.}
We focus on the contribution from a subregion $I\subset \scri$ with $u_i \le u \le u_f$.
As we have shown in section \ref{sec:renormalize}, the renormalized SP organized itself as a sum of different layers (equation \eqref{SPsum}).
It is convenient to rearrange these layers as
\be
\hat\Theta{}^R = \Theta_\text{C}+ \Theta_\text{rad} 
+ \sum_{p=1}^{\ell}\left( \Theta^\text{int}_{(p)} + \Theta^\text{int}_{(- p )}\right),
\ee
where
\begin{align}
\Theta^\text{int}_{(p)} := \int_I \q \;\J^i_{(\ell + p )} \delta A_{i ( \ell +1- p )} 
\end{align}
is the contribution of the ``intermediate potentials'' (present only when $\ell\geq1$, i.e. $D\geq8$), while $\Theta_\text{C}$ and $\Theta_\text{rad}$ are the Coulombic and the radiative contributions respectively:
\bea
\Theta_\text{C} = \int_I  \q \; \J^i_{(2\ell+1)} \delta A_{i (0)},
\qquad 
\Theta_\text{rad} = \int_I \q \;\J^i_{(\ell)} \delta A_{i (\ell +1)}.
\eea
The Coulombic and radiative contributions are common to all even dimensions $D\geq6$. These are the layers we analyze thoroughly in this paper.

In order to express the radiative component of the SP, one has to remember that the vanishing of the vector anomaly \eqref{vecanomaly} imposes\footnote{Recall also that in $D=6$ the same condition follows from the asymptotic Maxwell conditions.} that $ A_{u(\ell+1)}=-\J^u_{(\ell-1)}/(\ell+1)  =0$. Therefore, in absence of anomalies the radiative component is purely transverse and pairs the Maxwell news $\N^A:=\J^A_{(\ell)}$ to the radiative modes $\A_A:=A_{A(\ell+1)}$, hence its name:
\bea
\Theta_\text{rad} 
&=& \int_I \q\, \N^A \d \A_A.
\eea

The study of the Coulombic component is more subtle. 
Recalling our definitions of the soft potential \eqref{softpot} and of the charge aspect, $\sigma := \J^u_{(2\ell+1)}$, and making use of the Gauss law \eqref{eq:gaussk}, the Coulombic component can be cast in the form
\bea
\Theta_\text{C} &: =& \int_I \q \J^i_{(2\ell+1)} \d A_{i (0)}
= \int_I \q \J^A_{(2\ell+1)} \delta \alpha_{A(0)} + \int_I \q \J^i_{(2\ell+1)}\p_i \d \varphi \cr
& = & \oint_S \q  \langle \J^A_{(2\ell+1)}\rangle \delta \alpha_{A(0)} + 
\oint_{S}  \q \left[ \sigma \d \varphi \right]_i^f .
\eea
Here, we used the notation  $\left[ X \right]^f_i:= X(u_f)-X(u_i)$, as well as introduced the Fourier zero-mode
\be
\langle X  \rangle :=\int_{u_i}^{u_f} \q \,X(u) \rd u
\ee

This shows
that the charge aspect $\sigma_i=\sigma(u_i)$ (resp.$\sigma_f=\sigma(u_f)$) is canonically conjugated to 
$\varphi_i=\varphi(u_i)$ (resp. $\varphi_f=\varphi(u_f)$)
while the zero-mode of the current $\langle \J^A_{(2\ell+1)}\rangle$ is conjugated to  $\alpha_{A(0)}$.

It is convenient to introduce the charge aspect (semi-)sum, $\sigma^+$, and difference, $\sigma^-$:
\be 
\sigma^+:= \tfrac12(\sigma_f+\sigma_i),\qquad
\sigma^- := \sigma(u_f)-\sigma(u_i),
\ee
and similarly for the soft potential.
Using
\be
 \oint_{S}  \q \left[ \sigma \d \varphi \right]^f_i 
 = 
 \oint_S\q (\sigma^+ \delta \varphi^-  + 
\sigma^-  \d \varphi^+),
\ee
the Coulombic part of the soft potential can be finally written as
\be\label{Coulombsimple}
\Theta_\text{C} =
\oint_S \q  \Big( \langle \J^A_{(2\ell+1)}\rangle  \delta \alpha_{A(0)} +  \sigma^+ \delta \varphi^- +
\sigma^-  \d \varphi^+      \Big).
\ee

What is interesting in this formulation is that $\alpha_{A(0)}$, $\varphi^+$, and $\varphi^-$ have a clear meaning in terms of the leading gauge potential $A_{i(0)}$:
the soft potential difference $\varphi^- $ is equal to $ [\varphi]^{f}_{i} = \int^{u_f}_{u_i} A_{u(0)}$; the sum $\varphi^+$ is the electric component in the Hodge decomposition of $A^+_{A(0)}$  and since this expression does not depend on the retarded time explicitly, it does not enter $A_{u(0)}$. $\alpha_{A(0)}$ is the magnetic component in the Hodge decomposition of $A_{A(0)}$, which the asymptotic Maxwell conditions requires to be time independent (see \eqref{Fn12}).
Finally, and most importantly, using the results of section \ref{sec:newsdef}, $\sigma^-$ can be related to the (generalized) zero-mode of the radiative modes. 
This will be shown in the next two sections.

%----------------------------------------------------------------------------------------------------------

\subsection{$D=6$ }
\label{subsec:6D}

In $D=6$, which means $\ell=0$, the symplectic potential contains only two layers. Moreover, while $\A_A=A_{A(1)}$ is the radiative mode, the curvature $F_{AB (0)}$ vanishes by the vector anomaly (see equation \eqref{F6}). We thus get that $\alpha_{A(0)}=0$, in this case.
Hence, the Coulombic component simplifies further and reduces to the sole contribution \eqref{Coulombsimple}.
We also have in this case that the charge conservation is directly
 determined by the radiative zero modes,
 \be\label{Chargecons}
 \sigma^- 
=  \tfrac{1}{2}\big( D_AD^A- 2 \big)  D_B \langle \A^{B} \rangle.
 \ee
Given that the zero mode of $\A$ enters the Coulombic part of the potential, it is necessary to carefully disentangle the  zero mode contribution of $\A$ from its purely radiative component 
contained in $\p_u \A$.
To do so we introduce the Green's function $G$, inverse to $\p_u$ which satisfies the following 
\be
\p_u G(X)= X, \qquad \langle G(X) \rangle =0,\qquad 
 X = \frac{1}{( u_f-u_i )}\langle X\rangle +  G(\p_u X).
\label{eq:Gproperties}
\ee
These conditions determines $G$ uniquely and an explicit expression is given in appendix \ref{app:Green}.
Using it,  the radiative modes can be decomposed into their zero and non-zero Fourier modes,  $ \A=\frac{ 1}{(u_f-u_i)}\langle \A \rangle + G(\p_u \A)$. 
From the expression \eqref{News1} of the Maxwell news at $\ell=0$ we see that the equations of motion for the radiative more in $D=6$ are simply  
\be
 \p_u \A_{A} = \N_A.
\label{76}
 \ee
Adding the Coulombic component \eqref{Coulombsimple} and taking another (antisymmetrized) variation of the total SP, one obtains the total symplectic form $\hat\omega{}^R=\delta \hat \Theta{}^R$:
\bea
\hat\omega{}^R =
\oint_S\q \left( \d \sigma^+ \curlywedge \delta \varphi^-  +  
  D^A \d \langle\A_A \rangle \curlywedge  \frac12(D^2-2) \delta \varphi^+   
 \right) 
+  \int_I \q\, \d \N^A \curlywedge  \d \A_A.
\label{78}
\eea

To fully unravel the last component of the SP, we turn our attention to the Fourier analysis of the Maxwell news.
In order to have a finite energy flux $\int_\scri \N^2 <\infty$, the Maxwell news must be an $\mathrm{L}^2$ function of $u$.
This means that the Fourier transform of $\N^A$ exists, and we can define (here, $\Delta u := u_f - u_i$)
\be
\overline\N{}^A (u):=  \N^A(u) - n^A(0) := \frac{1}{\Delta u^{1/2}} \sum_{k\neq0} e^{\frac{2\pi i k}{\Delta u}u} n^A(k) 
\ee 
with $\langle{ \N^A } \rangle = \Delta u\, n^A(0)$ the zero Fourier mode of the Maxwell news. 
Now, thanks to \eqref{76}, and the defining properties of $G$, 
\begin{align}
\A_A(u) &= \frac{1}{\Delta u}\langle \A_A \rangle + \frac{u- u^+}{\Delta u} \langle \N_A \rangle+ \frac{1}{\Delta u^{1/2}}  \sum_{k\neq0} e^{\frac{2\pi i k}{\Delta u}u} a_A(k) \notag\\
& =: \frac{1}{\Delta u}\langle \A_A \rangle + \frac{u- u^+}{\Delta u} \langle \N_A \rangle+ \overline \A_A(u) ,
\end{align}
where $u^+ := \frac12(u_f+u_i)$ and 
\be
a_A(k):= \frac{\Delta u}{2\pi i k} n_A(k).
\ee
Notice that $\overline \A_A(u)$ is periodic in the interval $[u_i;u_f]$.
We see that allowing $\langle \N^A \rangle$ to be nonzero means that $\A_A(u)$ has a $u$-linear component. 
Thus, inserting the above expressions for $\A_A$ and $\N^A$  in the last term of \eqref{78}, we readily obtain
\be
\int_I \sqrt{q} \; \d \N^A \curlywedge \d \A_A  
 = \int_I \sqrt{q} \; \d \overline\N{}^A \curlywedge \d \overline \A_A 
+\oint_S \sqrt{q} \; \d \langle \N^A \rangle \curlywedge \d \left( \frac{\langle \A_A \rangle}{\Delta u} - \overline\A{}^+_A \right),
\label{eq:83}
\ee
where we used that\footnote{Recall that $\overline \A_A$ is periodic in $[u_i;u_f]$ and therefore $\overline\A{}^+_A = \overline \A{}_A(u_i) = \overline \A{}_A(u_f)$.\label{fn23}}
\be
\langle u   \overline \N{}_A\rangle = \Delta u \cdot \overline \A{}^+_A.
\ee
The first term on the right hand side of equation \eqref{eq:83} is the radiative contribution proper, involving only the oscillating modes of the radiative data, while the second term is a soft contribution that is usually overseen. 
We will discuss this contribution in the $I\to\scri$ limit below.

To summarize, the renormalized asymptotic symplectic form of electromagnetism in $D=6$ on $I\subset \scri$ is given by 
\begin{subequations}
\be
\omega^R = \omega^R_\text{rad} + \omega^R_\text{soft}
\ee\be
\omega^R_\text{rad} = \int_I \sqrt{q} \;  \d \overline\N{}^A \curlywedge \d \overline\A_A 
\ee\be
\omega^R_\text{soft}  = \oint_S \sqrt{q} \; 
\Big( 
\d \sigma^+ \curlywedge \d \varphi^- 
+  \d \mathcal S^A   \curlywedge  \d\langle \A_A \rangle 
+ \d  \overline\A{}^+_A  \curlywedge \d \langle \N^A \rangle 
 \Big),
\ee
\end{subequations}
where the {\it soft current}  $\cal S^A$ was introduced,
\be
\mathcal S^A := \frac12 D^A(D^2-2)  \varphi^+ + \frac{1}{\Delta u} \langle \N^A \rangle.
\ee

We see that the $6D$ theory contains one purely radiative canonical pair and three types of soft canonical pairs.

The first soft pair $\d \sigma^+ \curlywedge \d \varphi^-$ is purely Coulombic and  it pairs the charge aspect sum $\sigma^+ := \tfrac12(\sigma_f + \sigma_i)$ with the change  in the soft potential $\varphi^- := [\varphi]^{f}_{i} = \langle A_{u(0)}\rangle$. 
The second soft pair $\d \mathcal S^A \curlywedge \d \langle \A_A \rangle$ involves the zero-mode $\langle \A_A \rangle$ of the radiative field and contains itself two contributions. 
The  first one, involving $ \varphi^+ $, is related to charge conservation \eqref{Chargecons}. 
It plays a key role in the derivation of the soft theorems (see section \ref{sec:softvsgauge}). 
Notice that the component $A_{u(0)}$ of the gauge field does not enter $\varphi^+$. Rather, in $\varphi^+$ enters the electric component in a Hodge decomposition of $A_{A(0)}$ (see \eqref{softpot}). 
Here, in $D=6$, the purely magnetic part of $A_{A(0)}$, i.e. $\alpha_{A(0)}$, is zero as a consequence of the vanishing of the vector anomaly, equation \eqref{F6}.
The electric contribution of $\varphi^+$ is analogous to the one discussed in $D=4$ e.g. by \cite{Strominger:2017zoo}.
The important difference is that, in $D=4$, the radiative data and the analogue of the scalar $\varphi^+$ both live at leading order, while in $D=6$ and higher the radiative data and leading order data are neatly separated.

The second term in the second soft pair is new. Due to its $\Delta u^{-1}$ scaling, it is hard to individuate when working directly in the $I\to\scri$ limit. In particular, for this term to survive, one has to suppose that the product $\langle \A_A \rangle \langle \N^A \rangle$ diverges as $\Delta u$ in the limit $\Delta u \to \infty$. 
For the previous term to be finite and $\varphi^+$ to be of order 1, one needs to require that the purely magnetic part of $\langle \A_A\rangle$ in a Hodge decomposition is allowed to be of order $\Delta u$, and thus possibly divergent, even though  its purely electric part should be of order 1 because of its coupling to $\varphi^+$. 

Finally, the last canonical pair, $\d \overline \A_A^+ \curlywedge \d \langle\N^A\rangle$, is bound to vanish in the $I\to\scri$ limit, in which the radiative part of $\overline \A_A$ goes to zero\footnote{Recall that $\overline \A_A$ is periodic and hence $\overline \A{}^+_A = \overline \A_A(u_i) = \overline \A_A(u_f)$.} as $u_i\to-\infty$ and $u_f\to +\infty$. This corresponds to saying that there is no outgoing\footnote{Recall also that in this paper by $\scri$ we mean $\scri^+$. Analogous statements hold at $\scri^-$.} radiation in the asymptotic past and future of $\scri$.

%================================================

 \subsection{Higher dimensions}
 
 Now, we briefly turn to the higher dimensional case, i.e. $D\geq8$ (even) or equivalently $\ell\geq 1$, and focus on the soft contribution to the renormalized symplectic structure.
 In this case the key equation is \eqref{eq:gaussell+1}, i.e. 
\bea
\p_u^{\ell+1}\sigma 
= \frac{(-1)^{\ell}}{2^{(\ell+1)}} \frac{1}{ (\ell+1) !} \Big(
\Dk{2\ell+1}
\Dk{2\ell}
\cdots 
\Dk{\ell+1}\Big)](D_A \A^A).
\eea
where we also recall that
\be
\Dk{k}:= \Big( D^AD_A -(k+1)(2\ell+2-k)\Big).
\ee
From this, {\it assuming} $D^A\alpha_{A(k)}=0$ for $k\in\{\ell+1,\dots,2\ell+1\}$, one finds that the  soft potential sum $\varphi^+$ is conjugated to
\begin{align}
\sigma^- &= \int^{u_f}_{u_i} \rd u \, \p_u \sigma  
= \frac{(-1)^{\ell}}{2^{(\ell+1)}} \frac{1}{ (\ell+1) !} \Big(
\Dk{2\ell+1}
\Dk{2\ell}
\cdots 
\Dk{\ell+1}\Big)
\langle D^A\A_A \rangle_{(\ell)} 
\label{eq:sigma-}
\end{align}
where the ``generalized zero-mode'' is defined as\footnote{Thus $\langle\cdot\rangle_{(0)}= \langle \cdot \rangle$ of the previous section.}
\be
\langle D^A\A_A \rangle_{(\ell)} = \int^{u_f}_{u_i} \rd u_{\ell+1} \int^{u_{\ell+1}}_{u_i}
\rd u_{\ell} \cdots \int^{u_2}_{u_i} \rd u_1 D^A \A_A(u_1).
\ee

Notice that the neglected contributions proportional to $D^A\alpha_{A(k)}$ contain powers of the interval $(u_f-u_i)$ and therefore require a more subtle analysis.
Moreover, in these cases where $D\geq 8$, the intermediate potentials also contribute via $\Theta^\text{int}_{(p)}$. These contributions, once fully unraveled in terms of the free data, end up ``dressing'' the different contributions to the SP while also providing new terms involving $\delta \alpha_{A(k)}$. We do not attempt a full analysis here.

%==================================

%================================================

\subsection{Gauge modes, soft modes, and soft theorems\label{sec:softvsgauge}}

We conclude this section with an important remark.
First we recall that we have so far worked in the gauge 
$A_\Omega=0$. This means that we can express our symplectic potential in terms of a gauge invariant potential provided we perform the replacement
\be\label{firstgauge}
A_a \mapsto A_a - \p_a \int^\O_0 A_\O
\ee
It is important to note that 
the Coulombic contribution to the SP would have been missed completely, had we fully ``fixed the gauge'', as in
\be
 A_a \mapsto A_a - \p_a \int^u_{u_i} A_{u(0)},\qquad A_a \mapsto A_a - \p_a  \frac{1}{D^2}(D^A A^+_{A(0)}).
\ee
In fact the last two transformations would have set $\varphi^-$ and $\varphi^+$ to zero, respectively.\footnote{The soft contribution pairing the zero-modes of the news and radiative modes in a sense comes from the radiative contribution to the SP.}

The question remains, what the first  gauge fixing in \eqref{firstgauge} removes, since we have indeed employed it to solve the EoM.
Using the Gauss law, it is easy to see that the term that was missed leads to the ``total'' SP, which differs from the original potential by a corner term
\be
\Theta^R_\text{tot} = \Theta^R-\sum_{k=1}^{D-5} \frac1k \oint_S \left[ \J^u_{(D-5-k)}\delta A_{\O(k-1)}\right]^f_i.
\ee
This corner term is a  spacetime local expression in terms of the gauge field $A_a$ and can therefore be considered as part of the (finite) corner ambiguity $\hat\ba{}^{\O u}$.

This finite corner ambiguity has the peculiar property that the Noether charges of section \ref{sec:generators},
 associated to $\Theta^R_\text{tot}$ are {\it independent} of the way the gauge parameter is extended off of $\scri$, i.e. they are not layered.\footnote{In \cite{hopfmuller2018null}, a similar criterion was used by two of us to fix analogous ambiguities in the gravitational context.}
Indeed, using \eqref{eq:renorm_hamiltonian}, one readily finds
\be
\hat Q^{\text{tot},R}_\epsilon = \hat Q^R_\epsilon - \sum_{k=1}^{D-5} \oint_S \sqrt{q} \J^u_{(D-5-k)}\epsilon_{(k)} =  \oint_S \sqrt{q} \sigma \epsilon_{(0)}.
\ee
From this formula, it is clear that the Hamiltonian generator for a generic gauge transformation is
\be
\hat H^{\text{tot},R}_\epsilon = \left[\hat Q^{\text{tot},R}_\epsilon \right]^f_i = \oint_S \sqrt{q} \,\big( \sigma^+ \epsilon^-_{(0)} + \sigma^- \epsilon^+_{(0)}\big).
\ee

Let us compare this Hamiltonian generator to the results of \cite{Kapec:2014zla}. There, starting from the QED soft theorem in dimensions $D=6+2\ell$, the authors derive the charge expression whose Ward identity encodes the soft theorem. They then fix the classical Poisson brackets, or equivalently the symplectic form, by demanding that the charge expression generate the correct gauge transformations of the gauge field $A_{A(0)}$. Here, we took a different route. We determined the symplectic form using the covariant Hamiltonian formalism and our renormalization procedure, and derived the charge from the symplectic form rather than deriving the symplectic form from the charge. 

The charge expression of \cite{Kapec:2014zla} coincides with our $\hat H^{\text{tot},R}_\epsilon$, for $u_i \rightarrow - \infty$, $u_f \rightarrow + \infty$, and under the assumption made in \cite{Kapec:2014zla} that $\sigma(u_f) = 0$. That is, using \eqref{Chargecons},  in $D=6$ we find
\be
\hat H^{\text{tot},R}_\epsilon \to 
\oint_S \sqrt q \,  \epsilon_{(0)}\sigma_i = 
-\frac12
\oint_S \sqrt q \, \epsilon_{(0)} (D^AD_A-2) \langle D_A\A^A\rangle
\ee

In higher dimensions, the correct generalization is obtained through equations \eqref{eq:sigma-}, and also coincides with the results of \cite{Kapec:2014zla}  
\begin{align}
\hat H^{\text{tot},R}_\epsilon \to &
\oint_S \sqrt q \,  \epsilon_{(0)}\sigma_i \notag\\&= 
-\frac{(-1)^{\ell}}{2^{(\ell+1)}} \frac{1}{ (\ell+1) !}
\oint_S \sqrt q \, \epsilon_{(0)} \left[
\Dk{2\ell+1}
\Dk{2\ell}
\cdots 
\Dk{\ell+p} \cdots 
\Dk{\ell+1}\right]
 \langle D_A\A^A\rangle_{(\ell)}
\end{align}

In particular, the ``soft-theorem charge'' is not the total radial electric field, which would a priori lead to divergent charges, but only the finite part of its Laurent series, which is the charge aspect $\sigma$. Recall also that the charge aspect is the Coulombic part of the radial electric field created by a bulk charge density. The agreement of the charge obtained from the renormalization procedure with the charge obtained from soft theorems supports the physical viability of the asymptotic renormalization procedure in gauge theories.

\section{Conclusions\label{sec:conclusions}}

In this work, we have presented the renormalization of the asymptotic symplectic potential for electromagnetism in even dimensions $D\geq6$ at null infinity. We have constructed in detail the renormalized symplectic potential and the corresponding charges under general asymptotic condition, { requiring $F_{ab}$ to be $(D-5)$-polyhomogeneous.}
{With a few obvious modifications our methods can be readily adapted to 4D, too.} 
We also have presented the derivation of the asymptotic solution in terms of free data, including the corner data. And we used this to express the symplectic potential entirely in terms of the free data.
The motivation of this work was to present the main ideas of asymptotic renormalization for electromagnetism before delving into a similar analysis for general relativity, which is the logical next step.

An avenue that needs to be revisited, now that we have allowed more general boundary conditions, is the possibility to define canonically the subleading charges \cite{Lysov2014PRL,Campiglia2016,LaddhaMitra2018,LaddhaSen2018}. 
Such charges  corresponds to divergent transformation of the field in the absence of renormalization \cite{Campiglia2016}. 
Also, it is clear that our analysis can be extended to include the analysis of the odd-dimensional case (see \cite{He:2019jjk} for a recent discussion about odd dimension).

{ Another intriguing feature of our analysis is the fact that by choosing the radial evolution to be along a null vector, we can present the asymptotic equations in a form 
that is independent of the signature of the radial slicing}. It is tempting to envision that the canonical analysis presented here could also be useful to revisit and maybe extend some of the result established for asymptotic AdS spaces. In particular one may wonder whether it is possible to have an asymptotic symmetry algebra in AdS that includes more general generators than the conformal algebra.

Finally, we would like to mention one last possible application of these results.  
Now that the finite and asymptotic (pre)symplectic potentials contain the same number of modes, it is possible to compare calculations at finite and asymptotic boundaries directly. 
In particular, it is now possible to investigate under which circumstances
the soft modes can be understood as asymptotic edge modes (cf. \cite{Donnelly:2016auv}, and \cite{Gomes:2019rgg,Riello:2019tad} for a different approach).

\textbf{Acknowledgments:} We would like to thank Hal Haggard for his help in finishing this work. We thank Glenn Barnich for useful discussions. FH would like to thank the organizers and attendees of the Gravity@Prague 2018 workshop, and Geoffrey Comp\`ere for discussions and his lectures there. { Finally, the authors would like to thank an anonymous referee who pushed us to clarify certain aspects of our renormalization algorithm in relation to holographic renormalization.} Research at Perimeter Institute is supported by the Government of Canada through the Department of Innovation, Science and Economic Development Canada and by the Province of Ontario through the Ministry of Research, Innovation and Science. FH acknowledges a Vanier Canada Graduate Scholarship.

%==============================================================
%==============================================================
\appendix

%==============================================================

\section{Radial equations}\label{app:radial}

A function $P$ is said to be polyhomogeneous \cite{Polyhomogeneous}, if it has an asymptotic expansion around $\Omega=0$ of  the form
\be
P(\Omega) = \sum_{p,q\geq0} P_{p,q} \O^p (\ln \O)^q,
\ee
where for each $p$  only a finitely many $P_{p,q}$ are non-zero.
We introduce the space of $n$-polyhomogenous functions $C^\text{poly}_{n}$ as the space of functions $Y$of the form
\be
Y(\O) = \sum_{k=0}^n \O^k Y_{(k)}  + \O^{n+1} P(\O)
\ee
with $P$ polyhomogeneous. This also means that when $Y\in C^\text{poly}_{n} $ then
\be
\po^nY = n! Y_{(n)} + \Omega P(\Omega).
\label{eq:defF}
\ee
Notice the following crucial properties of $Y\in C_n^\text{poly}$:
\be
\int_0^\O Y \in C^\text{poly}_{n+1} 
\qquad \text{and}\qquad
\O\p_\O Y \in C^\text{poly}_n.
\ee

Consider now an equation of the form
\begin{align}\label{maineq}
 (n - \O \po ) X = Y,
\end{align}
for $Y\in C_{n}^\text{poly}$ and $n\geq 1$. 
The main result of interest to us is if the source is $n$-polyhomogeneous $Y\in C^\text{poly}_{n} $ then
\be
(X- Y_{(n)}\O^n \ln\O) \in  C^\text{poly}_{n}.
\ee 
We refer to $Y_{(n)}$ as the {\it anomaly} of this equation. It appears as a logarithmic counterterm in our renormalization procedure.

We also introduce a renormalized version of $X$ given by 
\be
X^R : = X - C_n(Y) 
\ee
with counterterm
\be
C_n(Y):=  \sum_{p=1}^{n} \frac{(n-p)!}{n!} \O^{p-1} {\partial^{p-1}_\Omega Y} - \O^n \ln\Omega \,Y_{(n)}.
\label{Cc}
\ee
One also shows that this renormalized element is simply  equal to the following combination:
\be\label{Xexpd}
X^R(\Omega) =
\Omega^n \left(\frac{1}{n!} \po^n X + \ln \O Y_{(n)} \right).
\ee
and that it is determined by $Y$ up to a constant term:
\be\label{Xexp}
X^R(\Omega) = \Omega^n \left( X_{(n)} -  \frac{1}{n!}
 \int_{0}^\O P(\omega) \rd \omega\right).  
\ee
where $X_{(n)}$ is a \emph{free} integration constant which appears as the limit 
when $\O\to 0$ of $\O^{-n}X^R$.
{From this last expression and the properties discussed above it is clear that
\be
\Omega^{-n}X^R \in C^\text{poly}_0.
\ee}
{ The differential analogue of \eqref{Xexp} is the manifestly anomaly-free radial equation
\be
(n-\O\po)X^R = \frac{1}{n!}\Omega^{n+1} P.
\ee}

As a last remark, we provide the expressions for the counterterm $C_n(Y)$.
Since $C_n$ is a linear function, it is enough to evaluate it on the monomials.
One finds that
\begin{align}\label{RC}
C_n(\O^k) = 
\begin{cases}
\frac{\Omega^k}{n-k} & \mbox{if } k<n\\
-\O^n \ln\O  + \O^n  H_n   & \mbox{if } k=n\\
\frac{\O^k}{k-n} \left[ \binom{k}{n}-1\right]  & \mbox{if } k>n 
\end{cases}
\end{align}}
where $H_n = \sum_{p=1}^n p^{-1}$ and $\binom{k}{n}$ is the binomial coefficient.

For the renormalization purpose one could also use a truncated renormalization scheme 
denoted $C^0_n$ obtains by truncating the $Y$ Laurent series.
The evaluation of 
 $C^0_n$ on monomials is given by
{
\begin{align}
C^0_n(\O^k) = 
\begin{cases}
\frac{\Omega^k}{n-k} & \mbox{if } k<n\\
-\O^n \ln\O  & \mbox{if } k=n\\
0  & \mbox{if } k>n 
\end{cases}
\end{align}
}

%---------------------------------------------------------------------------------------------------------------------------

\subsection{Proof}\label{anomaly3}

Let us first establish (\ref{Xexpd}).
 It is easy to check that $\po^k (n - \O \po) X = \big((n-k) - \O\po \big) (\po^k X)$,
Using these equations we can evaluate the following difference as 
a sum
\bea
X - \frac{1}{n!} \O^n \po^n X 
&=& 
\frac{1}{n!} \sum_{k=0}^{n-1} \Big( (n - k)!\ \O^k \po^k X - (n-(k+1))!\ \O^{k+1} \po^{k+1} X\Big)\cr
&=& 
\frac{1}{n!} \sum_{k=0}^{n-1} (n-(k+1))! \O^k \Big( (n - k)  - \O \po\Big) \po^k X\cr
&=& \frac{1}{n!}\sum_{k=0}^{n-1} (n-(k+1))! \O^k \po^k Y.
\eea
This establishes the  identity
\begin{align}
X - \frac{1}{n!} \O^n \po^n X = \sum_{k=1}^{n} \frac{(n-(k+1))!}{n!}\O^k  \po^k Y,
\end{align}
which is valid at any  $\O$. 
From this we can evaluate the renormalized $X$ as 
\bea
X^R
= \O^n \left( \frac{1}{n!}  \po^n X +  \ln\Omega Y_{(n)} \right).
\eea
This establishes the first main result  (\ref{Xexpd}).

	In order to prove \eqref{Xexp}, we use \eqref{Xexpd} together with the radial equation \eqref{maineq} for $X$ to derive a radial equation for $X^R$, which we can then solve. We have
	\begin{align}
		\O \po X^R =& \O \po \Big[ \O^n \left( \frac{1}{n!} \po^n X + \ln\O Y_{(n)} \right) \Big]\notag\\
		=& n X^R + \frac{1}{n!} \O^{n+1} \po^{n+1} X + \O^n Y_{(n)}.
	\end{align}
  	The $n$th derivative of \eqref{maineq} reads $- \O \po^{n+1} X = \po^n Y$ such that we get
  	\begin{align}
  		\O \po X^R =& n X^R + \O^n \left( Y_{(n)} - \frac{1}{n!} \po^n Y \right).
  	\end{align}
  Using \eqref{eq:defF} one obtains
  \begin{align}
  		(n - \O \po) X^R =& \frac{1}{n!} \O^{n+1} P \qquad \Leftrightarrow \qquad \po \big(\O^{-n} X^R\big) = - \frac{1}{n!} P,
  \end{align}
which leads to \eqref{Xexp} when integrated.

Finally, the proof of \eqref{RC} follows from a direct computation of the counterterm for $Y=\Omega^z$ where $z \in \mathbb{C}\backslash \mathbb{Z}$ is a complex power. Using the definition of \eqref{Cc} one can evaluate:
\be
C_n(\Omega^z)=\Omega^z \left(\frac{}{} \sum_{p=1}^n \frac{ (n-p)!\Gamma({z+1}) }{n!\Gamma(z-p+2)}\right) 
=\frac{\Omega^z}{z-n}\left(\frac{z(z-1)\cdots (z-n+1)}{n!}- 1\right).
\ee
which is manifestly analytic in $z$, with no pole at $z=n$.
By taking the limits $z\to k \in \mathbb{Z}^+ $ we find the stated result.
In particular, in the limit $z\to n$, we have 
\be
\frac{\Omega^{(n+\epsilon)}}{\epsilon}
\Big(\left(1+\tfrac{\epsilon}{n}\right)\left(1+\tfrac{\epsilon}{n-1}\right)\cdots \left(1+{\epsilon}\right)- 1 \Big)
= \Omega^{n}  H_n  + O(\epsilon),
\ee
to which one has to add the term $-\O^n\ln \O$ designed to cancel the anomalous logarithmic term in $X$.

%===============================================================

\section{Green's function}\label{app:Green}

In this appendix we define the Green's function $G$, which provides the inverse to $\p_u$. Since $\p_u$ possesses a zero mode, the Green's function is only determined after we impose a normalization condition.
We denote 
\be
\Delta u := u_f-u_i,\qquad \langle X\rangle := \int_{u_i}^{u_f} X(u)\rd u. 
\ee
We show that there exist a unique Green function $G: C(I)\to C(I)$, which  is such that 
\be\label{G1}
\p_u G(X) = X, \qquad \langle G(X)\rangle =0.
\ee
This Green's function is explicitly given by 
\be\label{G2}
G(X):= \int_{u_i}^u \frac{(v-u_i)}{(u_f-u_i)} X(v) \rd v - 
\int_{u}^{u_f} \frac{(u_f-v)}{(u_f-u_i)} X(v) \rd v,
\ee
and it also satisfy
\be\label{G3}
G(\p_u Y ) = Y -\frac{\langle Y\rangle}{\Delta u}. 
\ee
This Green's function is the solution of \eqref{G1} for $X=\delta(u-v)$ and its kernel is given by
\bea
G(u,v) &=& \theta(u-v)  \frac{(v-u_i)}{(u_f-u_i)} - \theta(v-u) \frac{(u_f-v)}{(u_f-u_i)}
\eea
This kernel is not skew, instead its symmetric and skew symmetric combinations $G_\text{a} = \tfrac12(G- G^t)$ and $G_\text{s}=\tfrac12(G+G^t)$ have the kernel
\be 
G_\text{a}(u,v) =\tfrac12 (\theta(u-v)- \theta(v-u)),\qquad
G_\text{s}(u,v) = \tfrac12 (v+u-( u_f+u_i))
\ee

{\bf Proof:}
To prove this statement we establish that (\ref{G1},\ref{G2},\ref{G3})
are equivalent.
First, it is easy to see that (\ref{G3}) is equivalent to  
(\ref{G1}) by simply taking $Y$ to be any  primitive of $X=\p_uY$.
We have that 
$\p_u G(X) = \p_uG(\p_uY)=  \p_u(Y -\frac{\langle Y\rangle}{\Delta u})=X.$ And the averaging condition $G(\langle\p_u Y\rangle)=0$ 
is obvious from the fact that $\langle 1 \rangle = \Delta u$.
We now show that  the explicit $G$ given in (\ref{G2}) satisfies (\ref{G3}). This follows from integration by parts:
\bea
G(\p_uY)&=& \int_{u_i}^u \frac{(v-u_i)}{(u_f-u_i)} \p_vY(v) \rd v - 
\int_{u}^{u_f} \frac{(u_f-v)}{(u_f-u_i)} \p_vY(v) \rd v\cr
&=& \left[ \frac{(v-u_i)}{(u_f-u_i)} Y(v)\right]_{u_i}^u -
\frac{1}{\Delta u }\int_{u_i}^u  Y(v) \rd v -
 \left[  \frac{(u_f-v)}{(u_f-u_i)} Y(v)\right]_{u}^{u_f}
- \frac{1}{\Delta u } \int_{u}^{u_f}  Y(v) \rd v\cr
&=& Y(u)- \frac{1}{\Delta u }\int_{u_i}^{u_f}  Y(v) \rd v.
\eea
Another derivation of the Green function is to start from the solution of 
(\ref{G1}) and derive  (\ref{G2}):
\bea
G(X)&=&
\int_{u_i}^u  X(v) \rd v - \frac{1}{\Delta u} \int_{u_i}^{u_f}  \left(\int_{u_i}^u  X(v) \rd v \right) \rd u \cr
&=& \int_{u_i}^u \rd v X(v)
- \frac{1}{\Delta u} \int_{u_i}^{u_f} \rd v \left(\int_{v}^{u_f} \rd u\right) X(v)\cr
&=&\int_{u_i}^u  X(v)\rd v
- \int_{u_i}^{u_f}  \frac{(u_f-v)}{(u_f-u_i)}X(v)\rd v \cr
&=& \int_{u_i}^u  \frac{(v-u_i)}{(u_f-u_i)} X(v)\rd v
- \int_{u}^{u_f}  \frac{(u_f-v)}{(u_f-u_i)}X(v)\rd v,
\eea
as promised.

\printbibliography

\end{document}